\documentstyle{article}

\newlength{\defaultparindent}
\newenvironment{Default Paragraph Font}{}{}
\input{tcilatex}

\begin{document}

\begin{center}
{\bf Unification of Weak and Gravitational Interactions Stemming from
Expansive Nondecelerative Universe Model}

Miroslav Sukenik$^{a}$, Jozef Sima$^{a}$ and Julius Vanko$^{b}$

$^{a}${\small Slovak Technical University, Dep. Inorg. Chem., Radlinskeho 9,
812 37 Bratislava, Slovakia}

$^{b}${\small Comenius University, Dep. Nucl. Physics, Mlynska dolina F1,
842 48 Bratislava, Slovakia}
\end{center}

{\bf Abstract}. There is a deep interrelationship of the General Theory of
Relativity and weak interactions in the model of Expansive Nondecelerative
Universe. This fact allows an independent determination of the mass of
vector bosons Z and W, as well as the time of separation of electromagnetic
and weak interactions.

In the early
stage of the Universe creation, i.e. in the lepton era an equilibrium of
protons and neutrons formation existed at the temperature about 10$^{9}$ - 10%
$^{10}$ K which corresponds to the energy of 1 MeV (a lower side of the range of weak interaction
energies). The amount of
neutrons was stabilized by processes including antineutrinos (eq.1) or neutrinos (eq.2) such as

$\nu +p^{+}\rightarrow n+e^{+}$ \qquad \qquad \qquad \qquad \qquad \qquad
\qquad (1)

$e^{-}+p^{+}\rightarrow n+\nu $ \qquad \qquad \qquad \qquad \qquad \qquad
\qquad (2)

The cross section $\sigma $related to the above processes can be expressed
[1, 2] as

$\sigma \cong \frac{g_{F}^{2}.E_{w}^{2}}{(\hbar .c)^{4}}$ \qquad \qquad
\qquad \qquad \qquad \qquad \qquad \qquad\ \ (3)

where $g_{F}$ is the Fermi constant (10$^{-62}$ J.m$^{3}$), $E_{w}$is the
energy of weak interactions that, based on (3), can be formulated by relation

$E_{w}\cong \frac{r.\hbar ^{2}.c^{2}}{g_{F}}$ \qquad \qquad \qquad \qquad
\qquad \qquad \qquad\ \ \ \ \ \ (4)

where $r$ represents the effective range of weak interactions. Stemming from
relation (4) it holds that in limiting case when

$r=\frac{\hbar }{m_{ZW}.c}$ \ \ \ \qquad \qquad \qquad \qquad \qquad \qquad
\qquad\ \ \ \ \ \ \ (5)

the maximum energy of weak interaction is given by

$E_{w}\cong m_{ZW}.c^{2}$ \qquad \qquad \qquad \qquad \qquad \qquad \qquad\
\ \ \ (6)

Relations (5) and (6) represent the Compton wavelength of the vector bosons
Z and W, and their energy, respectively. Equations (4), (5) and (6) lead to
expression for the mass of the bosons Z and W

$m_{ZW}^{2}\cong \frac{\hbar ^{3}}{g_{F}.c}\cong \left| 100GeV\right| ^{2}$
\qquad \qquad \qquad \qquad\ \ \ \ \ (7)

providing the value that is in good agreement with the known actual value.

For the density of gravitational energy $\varepsilon _{g}$ it follows from
the ENU model [3 - 5]

$\epsilon _{g}=-\frac{R.c^{4}}{8\pi .G}=-\frac{3m.c^{2}}{4\pi .a.r^{2}}$
\qquad \qquad \qquad \qquad \qquad\ \ \ (8)

where $\varepsilon _{g}$ is the gravitational energy density of a body with
the mass $m$ in the distance $r,R$ is the vector curvature, $a$ is the gauge
factor that reaches at present

$a\cong 10^{26}m$ \qquad \qquad \qquad \qquad \qquad \qquad \qquad\ \ \ \ \
\ \ (9)

As a starting point for unifying the gravitational and weak interactions,
the conditions in which the weak interaction energy $E_{w}$ and the
gravitational energy $E_{g}$of a hypothetic black hole are of identical value

$E_{w}=\left| E_{g}\right| $ \qquad \qquad \qquad \qquad \qquad \qquad
\qquad\ \ \ \ \ \ (10)

can be chosen. Based on relations (4), (8) and (10) in such a case it holds

$\frac{r.\hbar ^{2}.c^{2}}{g_{F}}=\left| \int \epsilon _{g}dV\right| \approx 
\frac{m_{BH}.c^{2}.r}{a}$ \qquad \qquad \qquad\ \ \ \ (11)

where $m_{BH}$ is the mass of a black hole and $r$ is the range of weak
interaction. It follows from (11) that

$m_{BH}\cong \frac{a.\hbar ^{2}}{g_{F}}$ \qquad \qquad \qquad \qquad \qquad
\qquad \qquad\ \ \ (12)

The above relation manifests that the mass of a black hole depends on the
gauge factor, i.e. it is increasing with time. On the other hand, the black
hole mass may not be lower than the Planck mass $m_{Pc}$

$m_{BH}\geq m_{Pc}$ \qquad \qquad \qquad \qquad \qquad \qquad \qquad\ \ \
(13)

that approximates

$m_{Pc}=\left( \frac{\hbar .c}{G}\right) ^{1/2}\cong 10^{19}GeV$ \qquad
\qquad \qquad\ \ \ \ (14)

The gravitational radius $l_{Pc}$ of a black hole having the minimum mass $%
m_{Pc}$ is

$l_{Pc}=\left( \frac{G.\hbar }{c^{3}}\right) ^{1/2}\cong 10^{-35}m$ \qquad
\qquad \qquad\ \ \ \ \ \ \ \ (15)

If there is a mutual relationship of the gravitational and weak
interactions, there had to be a time $t_{x}$ corresponding to a gauge factor 
$a_{x}$ when $m_{BH}$ and $m_{Pc}$ were of identical value. It was the time
of weak interactions formation. In such a case it stems from (12) and (13)
that

$a_{x}\cong \frac{m_{Pc}.g_{F}}{\hbar ^{2}}\cong 10^{-2}m$ \qquad \qquad
\qquad \qquad \qquad\ (16)

and

$t_{x}\cong 10^{-10}s$ \qquad \qquad \qquad \qquad \qquad \qquad \qquad\ \ \
(17)

This is actually the time when, in accordance with the current knowledge,
electromagnetic and weak interactions were separated (it might be worth
mentioning that its value represents also typical duration of weak
interaction processes). In the time $t_{x}$ it had to hold

$\frac{m_{Pc}}{m_{ZW}}=\left( \frac{a_{x}}{l_{Pc}}\right) ^{1/2}$ \qquad
\qquad \qquad \qquad \qquad \qquad\ \ (18)

Substitution of (16) into (18) leads to (7) which means that the mass of the
vector bosons Z and W as well as the time of separation of the
electromagnetic and weak interactions are directly obtained, based on the
ENU model, in an independent way.

Conclusions:

1.The Vaidya metrics [6] based ENU model allowing to localize the
gravitational energy exhibits its capability to manifest some common
features of the gravitational and weak interactions.

2.The paper presents an independent mode of determination of the mass of
vector bosons Z and W, as well as the time of separation of the
electromagnetic and weak interactions.

3.The paper follows up our previous contributions showing the unity of the
fundamental physical interactions.

References

1. I. L. Rozentahl, Adv. Math. Phys. Astr. 31 (1986) 241

2.L.B. Okun, Leptons and Quarks, Nauka, Moscow, 1981

3.V. Skalsky, M. Sukenik, Astrophys. Space Sci., 236 (1991) 169

4.J. Sima, M. Sukenik, General Relativity and Quantum Cosmology, Preprint
gr-qc 9903090

5.M. Sukenik, J. Sima, General Relativity and Quantum Cosmology, Preprint
gr-qc 9911067

6.P.C. Vaidya: Proc. Indian Acad. Sci., A33 (1951) 264

\end{document}